\documentclass[pra,superscriptaddress,amsmath,amssymb,submission,copyright,creativecommons]{eptcs}
\providecommand{\event}{QPL 2016} 
\usepackage{breakurl}             
\usepackage{amsfonts}

\usepackage{amsthm}
\usepackage{graphicx}
\usepackage[all]{xy}
\usepackage{color}

\theoremstyle{plain}

\theoremstyle{definition}
\newtheorem{definition}{Definition}

\title{Geometric Quantization and Epistemically Restricted Theories: The Continuous Case}
\author{Ivan Contreras
\institute{Department of Mathematics\\
University of Illinois at Urbana-Champain\\
Illinois, USA}
\email{icontrer@illinois.edu}
\and
Ali Nabi Duman
\institute{Department of Mathematics and Statistics\\
King Fahd University of Petroleum and Minerals\\
Dhahran, Saudi Arabia}
\email{aliduman@kfupm.edu.sa}
}
\def\titlerunning{Geometric Quantization and Epistemically Restricted Theories}
\def\authorrunning{I. Contreras, \& A.N. Duman}
\begin{document}
\maketitle

\begin{abstract}
It is possible to reproduce the quantum features of quantum states, starting from a classical statistical theory and then limiting  the amount  of knowledge that  an agent can have about an individual system \cite{B, S1}.These are so called epistemic restrictions. Such restrictions have been recently in terms of the symplectic geometry of the corresponding classical theory \cite{S}. The purpose of this note is to describe, using this symplectic framework, how to obtain a $C^*$-algebraic formulation for the epistemically restricted theories. In the case of continuous variables, following the groupoid quantization recipe of E. Hawkins, we obtain a twisted group $C^*$-algebra which is the usual Moyal quantization of a Poisson vector space \cite{H}.
\end{abstract}

\section{Introduction}

In 2003, Spekkens introduced as an evidence of an epistemic view of quantum states \cite{S1} in order to conceptually clarify certain quantum phenomena that resist classical explanation. Indeed, the new theory contains many prominent quantum features including complementarity, no-cloning, no-broadcasting, teleportation, entanglement, Choi-Jamiolkowski isomorphism and many others.

The toy theory was next reformulated by Coecke, et al. \cite{CES} in the language of symmetric monoidal categories \cite{M}, with the objects of the category representing systems and the morphisms representing processes undergone by these systems. This formalism, which falls under the programme of Categorical Quantum Mechanics initiated by Abramsky and Coecke \cite{AC}, not only provides a consistent mathematical formulation of the toy theory but it also identifies in a precise way the structural difference -phase group- between stabilizer quantum mechanics and the toy theory. Namely, the structure of the phase group carries the main physical difference between these two theories: non-/locality.

Spekkens has recently generalized the toy theory to continuous and finite variables by positing an epistemic restriction on what kind of statistical distributions can be prepared in a classical system. This new theory is called \emph{epistricted theory}. In this approach, starting with a  classical ontological theory one constructs a statistical theory where an object is a statistical distribution over the physical state space. Here, the deterministic dynamics change the statistical distribution. Liouville mechanics, statistical theory of bits, statistical optics are some examples of theories obtained at this point of the construction. In the final step, one put forward a restriction on the knowledge of an agent about the statistical distribution of the system. This results in subtheories of quantum mechanics such as Gaussian epistricted mechanics (or epistemically-restricted Liouville mechanics), stabilizer subtheory for qutrits, subtheory of quantum optics. For the case bits, this recipe gives us the original Spekkens' toy theory which resembles the stabilizer subtheory of qubits.

The above mentioned epistemical restriction, which is called \emph{classical complementarity}, states that "the valid epistemic states are those wherein an agent knows the values of a set of variables that commute relative to the Poisson bracket and is maximally ignorant otherwise." In order to describe this epistemic restriction, Spekkens uses a symplectic geometric formalism for which a symplectic inner product is derived from the Poisson bracket on the pair of functionals over the phase space. Hence, he points out the possibility of considering symplectic geometry as a possible scheme for an axiomatization programme of Quantum Mechanics.

A symplectic manifold is also considered to be the phase space of a classical system in an interesting mathematical abstraction of quantization, namely geometric quantization \cite{AE, BW, W}.
The general premise of this approach for quantization is to construct quantum objects (in particular quantum states, phase spaces, etc) by using the geometry of the corresponding classical objects. It originally started as an attempt to extend known quantization procedures, such as Weyl quantization, and more generally, deformation quantization, to broader configurations and phase spaces. As an outcome, the quantum objects come equipped with natural algebraic structures, from which the algebraic structures for the classical objects can be naturally recovered (this procedure is usually called the semiclassical limit).

In this setting, the aim is to construct a Hilbert space and observables of the underlying quantum theory from a symplectic manifold in order to compare classical and quantum theories. Another quantization scheme which applies to any Poisson manifold is deformation quantization. According to Riefel's approach, a deformation quantization of a Poisson manifold is a continuous field of $C^*$-algebras \cite{R}. This $C^*$-algebraic formulation can be related to geometric quantization via symplectic groupoid approach of Eli Hawkins \cite{H}. Our objective in this paper is to investigate how the epistricted theories fit into mathematical methods of geometric and deformation quantization.

Finding $C^*$-algebraic counterpart of the epistricted theories has several benefits. For example, one can investigate these theories from the operator algebraic point of view. In this direction, the work on stabilizer formalism for operator quantum error correction \cite{P} and the other related results \cite{KLPL, BKK} can provide insight about the foundational properties of the epistricted theories, such as complementarity, mutually unbiased bases, contextuality etc.

From the perspective of Categorical Quantum Mechanics, Vicary \cite{V} showed that finite dimensional $C^*$ algebras correspond to the Frobenius algebras over the category of finite dimensional Hilbert spaces. This result gives us a high level categorical formalism to study the characteristics of the epistricted theories. One can for example work on the functoriality between the subcategories of Hilbert spaces corresponding to Quantum subtheories and the subcategories of sets and relations corresponding to epistricted theories.

The outline of this paper is as follows. We begin section 2 with a summary of the geometric quantization procedure. We then discuss epistricted theories of continuous variables and their correspondence in geometric quantization framework. We next briefly review Eli Hawkins' groupoid quantization recipe from which we obtain the usual Moyal quantization as a twisted group $C^*$-algebra from the geometric formulation of epistricted theories. We finally conclude that the resulting $C^*$-algebra contains phase-space formalism for quadrature subtheories.  We end the paper with the conclusion and discussions.

A full discussion containing the finite case, and the derivation of the Frobenius algebras corresponding to epistemically restricted theories will be part of a future publication.

\section{Continuous degrees of freedom}
In this section, the main objective is to provide a conceptual interpretation of geometric quantization formalism under the light of epistemically restricted theories. Hence, we first give an overview of a certain segment geometric quantization literature which we are going to utilise to make the connection with the epistricted theories.

\subsection{Geometric quantization}

Construction of quantum mechanics from the geometry of classical mechanical objects is the main focus of the geometric quantization. Towards this objective, the geometric properties of a classical system are considered, in particular, symplectic geometrical features, as we will see in the example below.
The main problem of geometric quantization can be phrased as follows:
\begin{itemize}
\item Given a symplectic manifold $(M, \Omega)$ modelling a classical mechanic system and its geometric properties, construct a Hilbert space $\mathcal H$ and a set of operators on $\mathcal H$ which give the quantum analogue of the classical system.
\end{itemize}
\subsubsection{The particular case: the WKB method}
Here we follow the flow of the lecture notes of Bates and Weinstein \cite{BW} as they start with a specific case of WKB method and generalize this system to other phase spaces. Finally, one can get the connection to algebraic quantization (deformation quantization) via symplectic groupoid quantization. Thus, our final aim is to obtain a $C^*$-algebra structure which contains the epistricted theory.

On the other hand, the basic WKB picture stated in a symplectic geometric formalism is sufficient to cover the epistricted theories. WKB approximate solution of time-independent Schrodinger equation is $\phi=e^{iS/\hbar}$ where $S$ is a function satisfying Hamilton-Jacobi equation $H(x,\partial S/\partial x)=E$. 

Now, in more geometric terms, this solution can be realized as the lagrangian sub-manifold of the level set $H^{-1}(E)$. For a "semi-classical" approximation, one considers the transport equation
$$a\triangle S+2\sum \frac{\partial a}{ \partial x_j}\frac{\partial S}{\partial x_j}=0 $$
where $a$ is a function on $\mathbb{R}^n$. After multiplying both sides by $a$, one can find that the divergence of $a^2\nabla S$ is zero. By considering this condition on the lagrangian manifold $L=im(dS)$, we can deduce $\mathfrak{L}_{X^{(x)}_H}(a^2|dx|)=0$, where $X^{(x)}_H$ is the projection of the vector field $$X_{H|L}=\sum_{j}\frac{\partial S}{ \partial x_j}\frac{\partial}{\partial q_j}-\frac{\partial V}{ \partial q_j}\frac{\partial}{\partial p_j}$$ onto $\mathbb{R}^n$ where the hamiltonian $H$ is $H(q,p)=\sum p_i^2/2+V(q)$ and $|dx|=|dx_1\wedge \ldots \wedge dx_n|$ is the canonical density on $\mathbb{R}^n$. Since $X_H$ is tangent to $L$ by the Jacobi-Hamilton theorem and since Lie derivative is invariant under diffeomorphism, the equation $\mathfrak{L}_{X^{(x)}_H}(a^2|dx|)=0$ implies that the pullback $\pi^*(a^2|dx|)$ is invariant under flow $X_H$, where $\pi:T^*\mathbb{R}^n\rightarrow L$ is the projection onto $L$.

As a result, a geometric semi-classical state is defined as a lagrangian manifold $L$ of $\mathbb{R}^{2n}$ equipped with a function $a$ (half-density). This state is stationary when $L$ lies in the level set of the hamiltonian and $a$ is invariant under its flow and the transformations correspond to Hamiltonians.
Table 1 shows the correspondence between semi-classical (geometric) and quantum (algebraic) objects in this specific case.
\subsubsection{The general case}
If we follow Dirac's approach for quantization, we aim for a linear map $P\to \hat P_{\mathcal H}$, between a Poisson algebra $P$ and the algebra of operators on a (pre)-Hilbert space $P_{\mathcal H}$, satisfying compatibility conditions with respect to the Poisson bracket and the commutator on operators, as well as a compatibility between the complex conjugation on the left hand side and the adjunction on the right hand side. 
The objective in geometric quantization is to provide the Hilbert space $\mathcal H$ and the (pre)quantization map. This is achieved by constructing:
\begin{enumerate}
\item A (prequantum) line bundle: The prequantization procedure can be constructed explicitly in the case in which the Poisson algebra is the algebra of functions of a cotangent bundle $T^*N$. For more general Poisson algebras, the Hilbert space $\mathcal H$ is the space of sections of a complex line bundle over the Poisson manifold $M$, with an Hermitian structure and a compatible Hermitian connection.
\item A polarization: A prequantization is called quantization if in addition to the previously mentioned compatibility conditions, the image of a complete set of functions in $P$ is also a complete set of operators in $P\to \hat P_{\mathcal H}$.
It is not hard to note that the Hilbert prequantization space (by using the complex line bundle) is too large for this last compatibility condition to hold. The way out of this problem is to choose only half of the coordinates from the classical system to write down the quantized functions. This is inspired from the way we can write the wave functions in quantum mechanics by choosing half of the coordinates of the classical phase space. This procedure is called polarization.
\end{enumerate}

\subsubsection{Some words on symplectic geometry}
To be consistent with the formalism of epistricted theories, we define the lagrangian condition for a symplectic vector space as follows: For a symplectic vector space $(V,\omega)$, $\omega$\emph{-orthogonal} to a subspace $W\subset V$ is defined as the set $W^{\bot}=\{x\in V: \omega(x,y)=0, \forall y\in W \}$. A subspace is called \emph{isotropic} if it is contained in its orthogonal. Any self orthogonal subspace, i.e. $W=W^{\bot}$, is called  \emph{lagrangian}. As a result of non-degeneracy of $\omega$, one can obtain $\dim W= \frac{1}{2}\dim V $.

\begin{table*}[t]
  \centering
\begin{tabular}{|l|p{6cm}|p{6cm}|}
  \hline
  Object & Semi-classical (geometric) version& Quantum (algebraic) version\\ \hline \hline
  phase space & $(\mathbb{R}^{2n},\omega)$ & Hilbert space $\mathfrak{H}_{\mathbb{R}^{2n}}$ \\ \hline
  state & lagrangian submanifold of $\mathbb{R}^{2n}$ with half-density & half-density on $\mathbb{R}^{n}$ \\ \hline
  transformations & hamiltonian $H$ on $\mathbb{R}^{2n}$ & operator $\hat{H}$ on smooth half densities \\ \hline
  stationary state & lagrangian submanifold in level set of $H$  with invariant half-density & eigenvector of $\hat{H}$ \\
  \hline
\end{tabular}
\caption{Correspondence between classical and quantum objects}
  \label{tab:1}
\end{table*}

\subsection{Quadrature Epistricted Theories}

We now introduce the quadrature epistricted theories for continuous variables \cite{S}. The epistemic restrictions on classical variables are adopted from the condition of the joint measurability of quantum observables. A set of variables are \emph{jointly knowable} if and only if it is commuting with respect to the Poisson bracket. The other restriction besides joint knowability is that an agent can know only the variables which are linear combination of the position and momentum variables.

If we start with the phase space $\Omega=\mathbb{R}^{2n}$ where a point is denoted by $\mathbf{m}=(\mathbf{p}_1,\mathbf{q}_1,\ldots,\mathbf{p}_n,\mathbf{q}_n)$, epistemic restrictions imply that the functionals $f:\Omega \to \mathbb{R}$ are of the form $$f=\mathbf{a_1}q_1+\mathbf{b_1}p_1+\ldots + \mathbf{a_n}q_n + \mathbf{b_n}p_n+ \mathbf{c}$$ where $\mathbf{a_1}, \mathbf{b_1},\ldots, \mathbf{a_n},\mathbf{b_n},\mathbf{c}\in \mathbb{R}$ and $p_i(\mathbf{m})=\mathbf{p_i}$ and $q_i(\mathbf{m})=\mathbf{q_i}$ are functionals associated with momentum and position , respectively. Hence, each functional $f$ is associated with a vector $\mathbf{f}=(\mathbf{a_1}, \mathbf{b_1},\ldots, \mathbf{a_n},\mathbf{b_n})$. It is not hard to show that the value of the Poisson bracket over the phase space is uniform and equal to the symplectic inner product: $$[f,g]_{PB}(\mathbf{m})=\sum_{i=1}^n(\frac{\partial f}{\partial q_i}\frac{\partial g}{\partial p_i}- \frac{\partial g}{\partial q_i}\frac{\partial f}{\partial p_i})(\mathbf{m})=\langle \mathbf{f},\mathbf{g}\rangle$$
where
$$\langle \mathbf{f},\mathbf{g} \rangle=\mathbf{f}^T J \mathbf{g}$$
and $J$ is the skew symmetric $2n\times 2n$ matrix with components $J_{ij}=\delta_{i,j+1}-\delta_{i+1,j}$. Hence, the vector space $\Omega$ becomes a symplectic vector space with the symplectic inner product $\omega = \langle \cdot, \cdot \rangle$. This allows us to give the geometric presentation of the quadrature variables.

The only set of variables jointly knowable are the ones that are Poisson commuting. In symplectic geometry, this set corresponds to the subspace $V$ of vectors whose symplectic inner product vanish, i.e. $\forall \mathbf{f}, \mathbf{g}\in V$  $ \langle \mathbf{f}, \mathbf{g} \rangle = 0 $. For a $2n$-dimensional phase space, the maximum possible dimension of such a $V$ is $n$. Such a maximal space is a Lagrangian space as defined above and it corresponds to the maximal possible knowledge an agent can have. In order to specify an epistemic state one should also set the values of the variables on $V$. The linear functional $v$ acting on a quadrature functional corresponds to the set of vectors in $\mathbf{v}\in V$ which is determined via $v(f)=\mathbf{f}^T \mathbf{v} $. That is,for every vector $\mathbf{v}\in V$ we obtain distinct value assignment.

In summary, a pure state in the epistricted theory consists of a Lagrangian subspace $V \in \mathbb{R}^{2n} $ and a valuation functional $v: \mathbb{R}^{2n} \to \mathbb{R}$. In geometric quantization, the half density function can be regarded as this valuation function.

On the other hand, the valid transformations are the symplectic transformations which maps the quadrature variables to itself. These transformations map a phase space vector $\mathbf{m}\in \Omega$ to $S\mathbf{m} + \mathbf{a}$ where $\mathbf{a}$ is a displacement vector and $S$ is $2n\times 2n $ symplectic matrix. The group formed by these transformations is called the \emph{affine symplectic group}, which is subgroup of the hamiltonian symplectomorphism group. Thus, each of these transformations can be obtained from a hamiltonian. Finally, the sharp measurements are parametrized by Poisson commuting sets of quadrature variables (isotropic subspaces $V$) and the outcomes are indexed by the vectors in $V$.

We summarize the correspondence between geometric quantization and epistricted theories in Table 2.

\begin{table*}[t]
  \centering
\begin{tabular}{|l|p{6cm}|p{6cm}|}
  \hline
  Object & Semi-classical version in quantization& Epistricted theories\\ \hline \hline
  phase space & $(\mathbb{R}^{2n},\omega)$ & $(\mathbb{R}^{2n},\omega)$ \\ \hline
  state & lagrangian submanifold of $\mathbb{R}^{2n}$ with half-density $a:\mathbb{R}^{2n}\to \mathbb{R}$& lagrangian subspace with a valuation function $v:\mathbb{R}^{2n}\to \mathbb{R}$ \\ \hline
  transformations & hamiltonian $H$ on $\mathbb{R}^{2n}$ & affine symplectic transformation \\
  \hline
\end{tabular}
\caption{Correspondence between geometric quantization and epistricted theories}
  \label{tab:1}
\end{table*}


\subsection{Groupoid Quantization}
The aim of this section is to point out that the epistricted theories can be quantized by a twisted polarized convolution $C^*$-algebra of a symplectic groupoid in the sense of E. Hawkins. The main idea in this method is to find a $C^*$-algebra which is approximated by a Poisson algebra of functions on a manifold. $C^*$-algebra quantization is mainly developed by the work of Rieffel where the quantization is stated as a continuous field of $C^*$-algebras $\{\mathcal{A}_{\hbar}\}$. Hawkins' construction gives a single algebra $\mathcal{A}_1$ by involving additional structures on the symplectic groupoid. 

In his approach, it is possible to reinterpret geometric quantization for a broader class of examples, coming from deformation quantization of Poisson algebras. This gives a rigorous treatment to the \emph{dictionary} strategy of Weinstein relating the symplectic category and its geometrically quantized counterpart \cite{BW}.

We start with the definition of symplectic groupoid.
A \emph{topological groupoid} $\Sigma$
is a groupoid object in the category of topological spaces, that is, $\Sigma$ consists of a space of $\Sigma_0$ of objects and a space $\Sigma_2$ of arrows, together with five continous structure maps:
\begin{itemize}
\item The source map $s:\Sigma_2\rightarrow \Sigma_0$ assigns to each arrow $g\in \Sigma_2$ its source $s(g).$

\item The target map $t:\Sigma_2\rightarrow \Sigma_0$ assigns to each arrow $g\in \Sigma_2$ its target $t(g)$. For two objects $x$, $y\in \Sigma_0$, one writes $g:x\rightarrow y$ to indicate that $g\in \Sigma_2$ is an arrow with $s(g)=x$ and $t(g)$.

\item If $g$ and $h$ are arrows with $s(h)=t(g)$, one can form their composition, denoted $hg$, with $s(hg)=s(g)$ and $t(hg)=t(h)$. If $g:x\rightarrow y$ and $h:y\rightarrow z$ then $hg$ is defined and $hg:x\rightarrow z.$ The composition map $m$ is defined by $m(h,g)=hg$.

\item The unit map $u:\Sigma_0\rightarrow \Sigma_2$ is a two sided unit for composition.

\item The involution map $-^* :\Sigma_2\rightarrow \Sigma_2$. Here, if $g:x\rightarrow y$ then $g^*:y\rightarrow x$ is two sided inverse for composition.
\end{itemize}

$\Sigma$ is said to be a groupoid over $\Sigma_0$.

\begin{definition} A \emph{Lie groupoid} is a topological groupoid $\Sigma$ where $\Sigma_0$ and $\Sigma_2$ are smooth manifolds, and such that the structure maps $s$, $t$, $m$, $u$ and $-^*$ are smooth. Moreover, $s$ and $t$ are required to be submersions so that the domain of $m$ is a smooth manifold.
\end{definition}

\begin{definition} A Lie groupoid $\Sigma$ is called a \emph{symplectic groupoid} if $\Sigma_2$  is a symplectic manifold with symplectic form $\omega$ and the graph multiplication relation $\mathfrak{m}=\{(xy,x,y):(x,y)\in \Sigma_2\}$ is a lagrangian submanifold of $\Sigma_2 \oplus \overline {\Sigma}_2 \oplus \overline{\Sigma}_2$, where $\overline{\Sigma}$ is the symplectic manifold $(\Sigma_2, -\omega)$.
\end{definition}
A natural example of a symplectic groupoid is the \emph{pair groupoid} of a symplectic manifold, in which $\Sigma_2=\Sigma_0\times \Sigma_0$ is equipped with the product symplectic structure, and the multiplication is given by the partial composition of relations: $m((a,b),(b,c))=(a,c)$.

As $\mathfrak{m}$ is lagrangian, one can find a unique Poisson structure on $\Sigma_0$ of a symplectic groupoid such that $s$ is a Poisson map and $t$ is anti-Poisson. Hence, we have the following definition.

\begin{definition} A symplectic groupoid $\Sigma$ is said to \emph{integrate} a Poisson manifold $\Omega$ if there exists a Poisson isomorphism from $\Sigma_0$ onto $\Omega$.
\end{definition}
For example, the pair groupoid $\Sigma_0\times \Sigma_0$ is an integration of the symplectic manifold $(\Sigma_0,\Omega)$, where the Poisson structure is the one induced by the symplectic structure $\omega$.

\subsubsection{Hawkins' approach}
 Here is the Hawkins' strategy for geometric quantization of a  manifold $\Omega$.For a detailed discussion, one can refer to \cite{H}.

\begin{itemize}
\item Construct a symplectic groupoid $\Sigma$ over $\Omega$.
\item Construct a prequantization $(\sigma, L, \nabla)$ of $\Sigma$.
\item Choose a symplectic groupoid polarization $P$ of $\Sigma$ which satisfies both symplectic and groupoid polarization.
\item Construct a "half form" bundle.
\item $\Omega$ is quantized by twisted, polarized convolution algebra $C^*_P(\Sigma, \sigma)$.

\end{itemize}


\subsubsection{The case of epistricted theories}
Our aim is to apply this construction to the case of episteimacally restricted theories with continuous variables. 
In the particular case of symplectic manifold $\Omega=\mathbb{R}^{2n}$ with symplectic form $\omega$, which is the context of the epistricted theories, we have the symplectic groupoid $\Omega \oplus \Omega^{*}$  integrating  the  symplectic vector space $\Omega$, where the multiplication is given by fiber addition on $\Omega^*=\{(p^1,p^2,\cdots , p^{2n})\}$, i.e. the symplectic integration comes equipped with Darboux coordinates.

More explicitly, $\hat{\omega}(u):v \mapsto \omega(u,v)$ gives a map $\hat{\omega}:\mathbb{R}^{2n} \to \mathbb{R}^{2n*}$. One obtains a symplectic structure
 $$\sigma((x,y),(z,w))=\omega(x,z)-\omega(y,w)$$ $$=\hat{\omega}(x-y)[\frac{z+w}{2}]-\hat{\omega}(z-w)[\frac{x+y}{2}].$$
 We identify $\mathbb{R}^{2n}\oplus \bar{\mathbb{R}}^{2n}$ with the cotangent bundle $T^*(\mathbb{R}^{2n})$ as follows: For the local coordinates of covectors $(u,\xi)$, $(v,\eta)$ in  $T^*(\mathbb{R}^{2n})$, the cotangent symplectic structure is $$\sigma^*((u,\xi),(v,\eta))=\xi(u)-\eta(v).$$ This gives us a symplectomorphism $\Phi:\mathbb{R}^{2n}\oplus \bar{\mathbb{R}}^{2n} \to T^*(\mathbb{R}^{2n})$ such that
 $$\Phi: (x,y)\mapsto (1/2(x+y),\hat{\omega}(x-y))$$ where $\Phi^*\sigma^*=\sigma$.  \footnote{ This example has also been studied by Hawkins (see example 6.2 \cite{H}).}

One can obtain the the Darboux coordinates $(q_1,\ldots, q_n, p_1,\ldots,p_n)$ of $T^*(\mathbb{R}^{2n})$ from the symplectomorphism $\Phi$. The projection of $T^*(\mathbb{R}^{2n})$ to $\mathbb{R}^{2n*}$ is a fibration of groupoids whose fibers are Lagragian. Thus this is a polarization of the symplectic groupoid given by
$$P=span\{\partial/\partial p_1,\ldots, \partial/\partial p_n \}$$
The symplectic potential which vanishes on $P$ can be chosen as $\theta_P= -p^i dq_i.$

 This polarization gives us the half-form pairing, which enables quantizable observables to be represented as operators on the Hilbert space $L^2(\mathbb{R}^{2n})$.Hence, this yields the correspondence between the kernels of operators on $L^2(\mathbb{R}^{2n})$ and Weyl symbols of these operators. This kernel $T$ of an operator $f$ is given by
 $$Tf(p,q)=C\int f(\frac{p+q}{2},\zeta)e^{i\zeta (q-p)/\hbar} d\zeta.$$

 The quantization procedure gives the twisted group algebra $C^*(\Omega^*, \sigma)$ where $\sigma:\Omega^* \times \Omega^* \rightarrow \mathbb{T}$, $\sigma(x, y)=e^{\frac{-i}{\{q,p \}}}$. This is the usual Moyal quantization of a Poisson vector space (see \cite{R2}).  In this setting, the observables corresponds to functions in classical phase-space and the Moyal product of functions is derived from the product of pair of observables. In  this case, the position and momentum operators correspond to the generators of the Heisenberg group and they are related to each other by a Fourier transform.

 We now show that quadrature quantum subtheories agree with the Moyal quantization, in the symplectic case. To be consistent with the formalism of \cite{S}, we work with projector valued measures (PVM) rather than Hermitian operators. PVMs are used in quantum information and quantum foundations to represent measurements as eigenvalues of Hermitian operators are operationally insignificant and serve as labels of outcomes. A \emph{projector-valued measure} with outcome set $K$ is a set of projectors $\{\Pi_k: k\in K \}$ such that $\Pi_k^2=\Pi_k$, $\forall k \in K$ and $\sum_k \Pi_k=I$. Hence the position (momentum) observable are the set of projectors onto position (momentum) eigenstates\footnote{In the continuous case one can also use Hermitian operators corresponding to the real valued functionals but the commutation relation of Hermitian operators does not have finite counterpart Therefore, Spekkens preferred to use PVMs in order to cover finite and continuous cases simultaneously.}:
 $$\mathcal{O}_q=\{\hat{\Pi}_q(\mathbf{q}): \mathbf{q}\in \mathbb{R}\}$$
 where $$\hat{\Pi}_q(\mathbf{q})=| \mathbf{q}\rangle _q \langle \mathbf{q}|.  $$

We now define a unitary representation of symplectic affine transformation to introduce the other quadrature observables. The projective unitary representation $\hat{V}$ of the symplectic group acting on the phase space $\Omega$ satisfies $\hat{V}(S)\hat{V}(S')=e^{i\phi}\hat{V}(S S')$ for every symplectic matrix $S:\Omega \to \Omega$ and where $e^{i\phi}$ is a phase factor. The action of this unitary is defined by the conjugation
$$\mathcal{V}(S)(\cdot)=\hat{V}(S)(\cdot)\hat{V}^{\dag}(S).$$
For single degree of freedom, let $S_f$ be the symplectic matrix that takes the position functional $q$ to a quadrature functional $f$ such that $S_f\mathbf{q}=\mathbf{f}$. Then the \emph{quadrature observable} associated with $f$ is defined as follows
 $$\mathcal{O}_f=\{\hat{\Pi}_f(\mathbf{f}): \mathbf{f}\in \mathbb{R}\}$$
 where
 $$\hat{\Pi}_f(\mathbf{f})=\mathcal{V}(S_f) (\hat{\Pi}_q(\mathbf{f})).$$

 For the $n$ degrees of freedom $\Omega=\mathbb{R}^{2n}$, the quadrature observable associated with $f$ is given by $$\mathcal{O}_f=\{\hat{\Pi}_f(\mathbf{f}): \mathbf{f}\in \mathbb{R}^{2n} \}$$ where $$\hat{\Pi}_f(\mathbf{f})=\mathcal{V}(S_f) (I\otimes \cdots\otimes \hat{\Pi}_{q_i}(\mathbf{f})\otimes \cdots \otimes I)$$ for $S_f\mathbf{q}_i=\mathbf{f}$. We also know that the set of quadrature observables $\{\mathcal{O}_{f_i}\}$ commute if and only if the corresponding functionals $\{f_i\}$ are Poisson-commuting (see \cite{S1}). Hence, the commuting set of quadrature observables can be labelled by isotropic subspaces of $\Omega$. This set defines a single quadrature observable $$\mathcal{O}_{V}=\{\hat{\Pi}_V(\mathbf{v}):\mathbf{v}\in V\}$$ where $$\hat{\Pi}_V(\mathbf{v})=\prod_{\mathbf{f}^{(i)}}\hat{\Pi}_{f^{(i)}}(f^{(i)}\mathbf{v}).$$

 On the other hand, in the geometric quantization procedure, any functional $f$ on $\Omega$ is mapped to a hermitian operator $\hat{f}$ in a prequantum Hilbert space which corresponds to the observable $\mathcal{O}_f=\{\hat{\Pi}_f(\mathbf{f}): \mathbf{f}\in \mathbb{R}^{2n}\}$.  Moreover,  the commutation relation for the observables in both quadrature subtheories and geometric quantization, is implied by the Poisson commutation relation of the classical observables. As the polarization is the commuting set of these hermitian operators, the state that is obtained after quantization is the operator $\hat{\Pi}_V(\mathbf{v})$.The choice of the vertical polarization for the groupoid $\Omega \oplus \Omega^*$ is the responsible of the correspondence between the two quantum states. The half-form pairing defined above can be computed in terms of the integral kernel of the projection operator $\hat{\Pi}_f$, which has Weyl symbol $f$. This establishes a correspondence between phase-space formalism and quantum mechanics, and Moyal product is deduced from this correspondence.

 In \cite{S1}, the operational equivalence quantum subtheories and epistricted theories is proven using Wigner representation which maps operators in Hilbert space to the functions in phase-space formulation of quantum mechanics. It is also well-known fact that the Wigner representation of an operator product is given by the Moyal product. As a result, geometric quantization with an appropriate choice of polarization is operationally equivalent to epistricted theories. We can also conlude that group algebra $C^*(H)=C^*(\Omega^*,\sigma)$, which is the Hilbert space considered as a group representation of the Heisenberg group $H$, contains the algebraic structure of quadrature subtheories.


\subsection{Functoriality}

The functoriality of geometric quantization is a delicate issue and it has been proven that the quantization that fits with the Schroedinger picture is in fact not functorial. There are several problems even before quantization, in particular, that the symplectic category is not quite a category, since the composition of Lagrangian correspondence is not in general well defined, and also that when it is defined, the composition is not continuous with the standard topology in the Lagrangian Grassmanian. The failure of geometric quantization to functorially represent Schroedinger's picture is given e.g. in Gotay's work \cite{Go}

However, the geometric quantization picture for symplectic groupoids turns out to be functorial with respect to the choices, i.e. the polarizations (the groupoid one), the half line bundle. The fact that the choices of polarizations are affine means that there is a higher structure for our C*-algebra quantization, namely, the objects are symplectic manifolds, 1-morphisms are Lagrangian polarizations and 2-morphisms are affine transformations between Lagrangian polarizations. These 2-morphims is reflected in C*-algebra automorphisms after quantization.

\section{Conclusion and further work}
We establish the relationship between geometric quantization and quadrature subtheories for the continous degrees of freedom. We conclude that the group algebra $C^*(H)$ for Heisenberg group $H$ contains the quadrature subtheories as a result of groupoid quantization procedure. One can use this fact to give operator algebraic approach to quantum optics.

This construction also suggests that there is a "geometric quantization" functor, from a subcategory of the category of groupoids to the category of $C^*$-algebras. Following \cite{CCH}, this corresponds to a functor from Frobenius algebras in the category \textbf{FRel} (Frobenius algebras in the category of sets and relations) to Frobenius algebras in the category of Hilbert spaces \textbf{FHilb}. The functor has to be defined in the subcategory of Frobenius algebras arising from symplectic groupoids, and the morphisms have to be adapted in order to obtain functoriality.

In  ongoing work, we investigate discrete degrees of freedom. The variables in this case are chosen from a finite field instead of real numbers. Even though Spekkens' original toy theory \cite{S} is contained in the case where finite field is $\mathbb{Z}/2$, we will first consider odd degrees freedom. The reason is that for $\Omega= (\mathbb{Z}/2)^n$ the discrete Wigner representation can take negative values and therefore the epistricted theory does not coincide with the quadrature subtheories \cite{S1}. "Quantization" schemes in finite degrees were not studied extensively but there are some result worth to check from the conceptual viewpoint of epistricted theories \cite{GH}. Our aim is to give a discrete version of groupoid quantization that also contains the functorial case of \cite{GH}. We conjecture that the resulting algebra is $C^*(H)$ for the finite Heisenberg group $H$. This finite $C^*$-algebra corresponds to a Frobenius structure via the construction of Vicary \cite{V}. Thus, one can study quantum phenomena such as complementarity in quadrature theories in this algebraic framework.

We end the paper with the sketch of our discrete quantization. The details will appear in the future work.
 \begin{itemize}
 \item We start with the special dagger frobenius algebra of epstricted theories, \textbf{Spek}, which is a subcategory of finite sets and relations, \textbf{FRel}.
 \item We then construct the groupoid $\mathcal{G}$ corresponding to \textbf{Spek} via the explicit connection in Heunen et. al.\cite{CCH}.
 \item We next obtain the pair groupoid from $\mathcal{G}$ and introduce the symplectic structure on it which is compatible with the pair groupoid. In this case, the each polarization corresponds to a lagrangian subspace in epistricted theories.
 \item We then apply geometric quantization procedure on the pair groupoid by considering the complex valued function space on the groupoid and using discrete fourier transform (integral kernel) defined by Gross \cite{G}.
 \item Finally, we end up with the finite dimensional $C^*$-algebra from which one can construct special dagger frobenius algebra over \textbf{FHilb} via \cite{V}.
 \end{itemize}

\section{Acknowledgements}
We thank to Chris Heunen for bringing the question of complementarity in Spekkens' toy theory to our attention. AND acknowledges King Fahd University of Petroleum and Minerals (KFUPM) for funding this work through project No. SR141007. IC was partially supported by the SNF grant PBZHP2-147294.

\nocite{*}
\bibliographystyle{eptcs}

\begin{thebibliography}{50}
\bibitem{AC}
S. ~Abramsky, B. ~Coecke,
\newblock \emph{A categorical semantics of quantum protocols},
\newblock In: Proceedings of the 19th Annual IEEE Symposium on Logic on Computer Science (LICS'04), 415-425. doi: \href{http://dx.doi.org/10.1109/LICS.2004.1319636}{10.1109/LICS.2004.1319636}.

\bibitem{AE}
S. T. ~Ali, T. ~Rudolph and M. Englis,
\newblock \emph{Quantization methods: a guide for physicist and analyst},
\newblock Rev. Math. Phys. \textbf{17} (2005), no. 4, 391-400. doi: \href{http://dx.doi.org/10.1142/S0129055X05002376}{10.1142/S0129055X05002376}.

\bibitem{BKK}
C. ~Beny, A. ~Kempf and D. W. ~Kribs,
\newblock \emph{Generalization of Quantum Error Correction via Heisenberg Picture},
\newblock Physical Review Letters \textbf{98} (2007), 100502. doi: \href{http://dx.doi.org/10.1103/PhysRevLett.98.100502}{10.1103/PhysRevLett.98.100502}.

\bibitem{BW}
S. ~Bates, A. ~Weinstein,
\newblock \emph{Lectures on Geometry of Quantization},
\newblock Berkeley Mathematics Lecture Notes, Vol. 8,(American Mathematical Society, Providence, RI, 1997).

\bibitem{B}
S. D. ~Bartlett, T. ~Rudolph and R. W. Spekkens,
\newblock \emph{Reconstruction of Gaussian quantum mechanics from Liouville mechanics with an epistemic restriction},
\newblock Physical Review A \textbf{86} (2012), no. 1, 012103. doi: \href{http://dx.doi.org/10.1103/PhysRevA.86.012103}{10.1103/PhysRevA.86.012103}.



\bibitem{CES}
B.~Coecke, B. ~Edwards and R. W. Spekkens,
\newblock \emph{Phase Groups and the Origin of Non-locality for Qubits},
\newblock Electronic Notes in Theoretical Computer Science \textbf{270} (2011), no. 2, 15--36. doi: \href{http://dx.doi.org/10.1016/j.entcs.2011.01.021}{10.1016/j.entcs.2011.01.021}.

\bibitem{CCH}
A.~Cattaneo, I. ~Contreras and C. ~Heunen,
\newblock \emph{Relative Frobenius Algebras are Groupoids},
\newblock Journal of Pure and Applied Algebra (2013), 217:114-124. doi: \href{http://dx.doi.org/10.1016/j.jpaa.2012.04.002}{10.1016/j.jpaa.2012.04.002}.

\bibitem{G}
D.~Gross,
\newblock \emph{Hudson's Theorem for finite-dimensional quantum systems},
\newblock Journal of Mathematical Physics \textbf{47} (2006), 122107.

\bibitem{Go}
M.J. ~Gotay,
\newblock \emph{Functorial Geometric Quantization and Van Hove's Theorem},
\newblock International Journal of Theoretical Physics \textbf{19} (1980), 139--161.

\bibitem{GH}
S.~Gurevich and R. ~Hadani,
\newblock \emph{Quantization of symplectic vector spaces over finite fields},
\newblock Journal of Symplectic Geometry \textbf{7} (2009), no. 4, 475--502. doi: \href{http://dx.doi.org/10.4310/JSG.2009.v7.n4.a4}{10.4310/JSG.2009.v7.n4.a4}.

\bibitem{GB-V}
J.M. ~Gracia-Bondia and J.C. Varilly,
\newblock \emph{From Geometric Quantization to Moyal Quantization},
\newblock Journal of Mathematical Physics \textbf{36}(2691)(1995),
\newblock arxiv:\href{http://arxiv.org/abs/9406170}{9406170}. doi: \href{http://dx.doi.org/10.1063/1.531059}{10.1063/1.531059}. doi: \href{http://dx.doi.org/10.1017/CBO9781139034807.008}{10.1017/CBO9781139034807.008}.

\bibitem{H}
E.~Hawkins,
\newblock \emph{A groupoid approach to quantization},
\newblock Journal of Symplectic Geometry \textbf{6} (2008), no. 1, 61--125. doi: \href{http://dx.doi.org/10.4310/JSG.2008.v6.n1.a4}{10.4310/JSG.2008.v6.n1.a4}.

\bibitem{KLPL}
D. W. ~Kribs, R. ~Laflamme, D. ~Poulin and M. ~Lesosky
\newblock \emph{Operator Quantum Error Correction},
\newblock Quantum Information \& Computation \textbf{6} (2006), issue 4, 382--399.

\bibitem{M}
S.~Maclane,
\newblock \emph{Categories for the Working Mathemaician},
\newblock Second Edition (Springer--Verlag, 2000).

\bibitem{P}
D. ~Poulin,
\newblock \emph{Stabilizer Formalism for Operator Quantum Error Correction},
\newblock Physical Review Letters \textbf{95} (2005), 230504. doi: \href{http://dx.doi.org/10.1103/PhysRevLett.95.230504}{10.1103/PhysRevLett.95.230504}.

\bibitem{R}
M. A. ~Rieffel,
\newblock \emph{Quantization and $C*$-algebras},
\newblock Contemporary Mathematics \textbf{167} (1994), 67--97.

\bibitem{R2}
M. A. ~Rieffel
\newblock \emph{Deformation quantization for actions of $\mathbb{R}^d$},
\newblock Mem. Am. Math. Soc. \textbf{106}(506) (1993).doi: \href{http://dx.doi.org/10.1090/memo/0506}{10.1090/memo/0506}.

\bibitem{S1}
R. W. ~Spekkens,
\newblock \emph{Evidence for the epistemic view of quantum states: A toy theory},
\newblock Physical Review A \textbf{75} (2007), no. 3, 032110. doi: \href{http://dx.doi.org/10.1103/PhysRevA.75.032110}{10.1103/PhysRevA.75.032110}.


\bibitem{S}
R. W. ~Spekkens,
\newblock \emph{Quasi-quantization: classical statistical theories with an epistemic restriction},
\newblock arxiv: \href{http://arxiv.org/abs/1409.5041}{1409.5041}, 2014. doi: \href{http://dx.doi.org/10.1007/978-94-017-7303-4\_4}{10.1007/978-94-017-7303-4\_4}.

\bibitem{V}
J. ~Vicary,
\newblock \emph{Categorical Formulation of Finite-Dimensional Quantum Algebras},
\newblock Communications in Mathematical Physics \textbf{304} (2011), no. 3, 765--796. doi: \href{http://dx.doi.org/10.1007/s00220-010-1138-0}{10.1007/s00220-010-1138-0}.



\bibitem{W}
N. M. J. ~Woodhouse,
\newblock \emph{Geometric Quantization},
\newblock (Oxford University Press, New York, 1992).






%

\end{thebibliography}

\end{document}